\documentclass[aps,
reprint,
showpacs,
superscriptaddress,
10pt,
nofootinbib, 
prd
]{revtex4-2}
\usepackage{amsmath,amsthm, amssymb,amsfonts,graphicx,bm,dcolumn,color}
\usepackage{mathrsfs}
\usepackage{textcomp}
\usepackage[T1]{fontenc}
\usepackage{lmodern}
\usepackage[shortlabels]{enumitem}
\usepackage[breaklinks=true]{hyperref}
\usepackage{xurl}
\usepackage{caption}
\usepackage{braket}
\usepackage{simpler-wick}
\usepackage{physics}
\usepackage[compat=1.0.0]{tikz-feynman}
\usepackage{mathtools, slashed}
\usepackage{verbatim}
\usepackage{stfloats}
\usepackage{ragged2e}
\newcommand{\Fcan}{\mathcal{F}}

\usepackage[normalem]{ulem}
\newtheorem{definition}{Definition}[section]
\newtheorem*{remark*}{Nonextensivity criterion}
\makeatletter
\def\@fpheader{\relax}
\makeatother
\makeatletter
\def\maketag@@@#1{\hbox{\m@th\normalfont\normalsize#1}} 
\makeatother

\begin{document}

\title{Nonadditivity in Quantum Field Theory:\\ Replica Energies, Scaling Filters, and the Renormalization Group
}
\author{Giacomo Santoni}
\email{giacomo.santoni@alumni.sns.it}
\affiliation{Physics Department, INFN Roma1, Piazzale A. Moro 2, Roma, I-00185, Italy}

\author{Francesco Scardino}
\email{francesco.scardino@uniroma1.it}
\affiliation{Physics Department, INFN Roma1, Piazzale A. Moro 2, Roma, I-00185, Italy}
\affiliation{Physics Department, Sapienza University, Piazzale A. Moro 2, Roma, I-00185, Italy}

\begin{abstract}
Extensive systems have a simple thermodynamic signature: the dimensionless canonical free energy scales homogeneously with the size of the system. We show that the failure of this scaling, measured by the replica energy ${\cal E}$, provides a useful bridge between statistical mechanics and quantum field theory. The associated differential operator $(1-\frac1d L\partial_L)$ removes the leading bulk contribution to $\Fcan=\beta F_{\rm can}=-\log Z$ and isolates the part that is sensitive to boundaries, topology, defects, long-range forces, or other sources of nonadditivity. In quantum field theory this thermodynamic idea has two closely related uses. For ordinary finite-volume or spherical partition functions, suitable higher-order versions of the same filter remove local counterterms and extract universal fixed-point data such as the central charge, the sphere free energy $F$, and the Euler anomaly coefficient $a$. For replica geometries with entangling defects, the same filtering principle gives the renormalized defect free energy. In $2+1$ dimensions, its $n\to1$ limit gives the entropic $F$-function, with the sign fixed below by the standard free-energy convention.
 We use this perspective to distinguish ordinary finite-size corrections, topology-dependent constants in gapped phases, subextensive fracton degeneracies, and genuinely nonextensive systems with long-range interactions such as self-gravitating thermal matter. Replica energy therefore offers a common thermodynamic language for additivity, defect free energies, and renormalization-group irreversibility.
\end{abstract}

\maketitle

\section{Introduction}
\label{sec:intro}

Thermodynamics rests on two closely related structural properties: \textit{extensivity}, implied by Euler homogeneity of the free energy in the thermodynamic variables, and \textit{additivity}, which allows the factorization of the free energy under a decomposition into macroscopic subsystems. While these properties hold for a broad class of short-range interacting systems in the thermodynamic limit, they can fail in controlled ways at finite size, in long-range interacting systems, and in phases whose low-energy sector carries global or subsystem constraints.
\par In this work we use a quantitative and field-theoretically natural diagnostic of these failures, based on the Helmholtz free energy $F_{\rm can}=-\beta^{-1}\log Z$ and its dimensionless form
\begin{align}
	\Fcan \equiv \beta F_{\rm can}=-\log Z,
\end{align}
and on the systematic bookkeeping of deviations from Euler homogeneity. More generally, on a compact Euclidean manifold $\mathcal M$ we use
$\Fcan_{\mathcal M}\equiv-\log|Z[\mathcal M]|$, without implying the presence
of a thermal circle.
\par The central objects are the replica energies ${\cal E}$ and ${\cal E}_i$  \cite{Latella2015ThermodynamicsON} introduced in Eq.~\eqref{eq:eulerviolation}, which measure, respectively, the violation of extensivity and the violation of additivity under shape deformations. The main conceptual point is that in quantum field theory (QFT) replica energies can be computed (or constrained) by exact Ward identities and renormalization group (RG) equations, and they act as filters that remove the leading extensive part of $\Fcan$ and isolate universal subextensive contributions.

\subsection{Main results}

For QFTs with no particle-number term we define the \textit{isotropic replica energy}
\begin{align}
\label{eq:intro_defE}
\beta{\cal E}(L,\beta) \equiv \Fcan(L,\beta)-\frac{1}{d}\,L\partial_L \Fcan(L,\beta),
\end{align}
the isotropic replica energy vanishes identically when the canonical free energy is extensive, i.e. $\Fcan\propto L^d$.

\par In Subsec. \ref{subsec:filter} we introduce the concept of \emph{filter} as a differential operator acting on $\Fcan$ that singles out the nonextensive contributions. We relate the filter to the replica trick for the calculation of the entanglement entropy.

\par In Subsec. \ref{subsec:dward} we provide an exact representation of ${\cal E}$ in terms of the dilatation Ward identity and RG data. In particular, using the operator identity for the trace of the improved energy-momentum tensor, one obtains the integrated-trace formula
\begin{align}
	\label{eq:intro_trace}
	\beta{\cal E}
	=
	\Fcan+\frac{\beta}{d}U
	-\frac{1}{d}\int_{\Sigma_L\times S^1_\beta} d^{d+1}x\,
	\langle{\Theta_\mu}^{\mu}\rangle,
\end{align}
and an equivalent RG form involving $\beta(g)\partial_g \Fcan$ and $m^2\nu(g)\partial_{m^2}\Fcan$.

\par In Subsec. \ref{subsec:liquid} we observe that, beyond relativistic QFT, the double filter also provides a sharp discriminator between constant (liquid) topological order and linear subextensive fracton scaling, which are otherwise both subextensive in $\Fcan$.

\par In Subsec. \ref{subsec:Ftheorem_precise} we review the results of Ref.~\cite{FEletter}. We consider the round three-sphere $S^3_R$ of radius $R$. We introduced the second-order double filter
\begin{align}
    \mathcal{D}_{\rm ren}^{(3)} \equiv \left(1-D\right)\left(1-\frac{1}{3}D\right)
\end{align}
where $D=R\partial_R$, and found that the function
\begin{align}
    F_{\cal E}(R) = \mathcal{D}_{\rm ren}^{(3)}\Fcan_{S^3}(R)
\end{align}
is scheme independent and locally decreasing near fixed points at order $O(g^2)$ under any relevant scalar deformation of a three-dimensional CFT. However, an exact calculation in a free massive scalar theory reveals that $F_{\cal E}(R)$ is not monotone along the renormalization group flow, and hence fails to define a monotone $F$-function \cite{Klebanov:2011, Pufu:2016}. In the present work we clarify the relation of this quantity with the $F$-function constructed in Ref. \cite{CasiniHuerta:2012}
\begin{align}
    F=RS_{EE}'(R)-S_{EE}(R)=-(1-D)S_{EE}(R)
\end{align}
from the entanglement entropy $S_{EE}(R) $. We note that while $F_{\cal E}(R)$ is constructed by applying a second-order filter to $\Fcan_{S^3}(R)=-\log |Z_{S^3}(R)|$, the function $F$ is defined by applying a \emph{first-order} filter to $S_{EE}(R)$. Purely extensive contributions are removed before the application of the filter as part of the replica trick.

\par In Subsec. \ref{subsec:atheorem_precise} we specialize to the case of the round four-sphere $S^4_R$ of radius $R$. Diffeomorphism invariance constrains the UV sensitive part of $\Fcan_{S^4}(R)$ to scale as $R^4$ and $R^2$. Similarly to the case of the round three-sphere studied in the companion paper \cite{FEletter}, this motivates the canonical second-order double filter
\begin{align}
	&\mathcal{D}_{\rm pow}^{(4)} \equiv \left(1-\frac{1}{2}D\right)\left(1-\frac{1}{4}D\right),
\end{align}
where $D=R\partial_R$. This filter removes the scheme-dependent power-law terms in the dimensionless free energy. The remaining divergence has the form
\begin{align}
	-4a\log\left(\Lambda R\right)
\end{align}
where $a$ is the Euler anomaly coefficient on the round sphere \cite{Minahan:2021_deformed_spheres}. We show that the function
\begin{align}
	{\cal A}_{\cal E}(R)=-D\mathcal{D}_{\rm pow}^{(4)}\Fcan_{S^4}(R)
\end{align}
extracts $4a$ at the RG fixed points and that, unlike what happens in the 3d case \cite{FEletter}, ${\cal A}_{\cal E}(R)$ is monotonically decreasing as a function of $mR$ and goes to $-\infty$ as $mR\to\infty$ for a free massive scalar field.

\subsection{Applications}

The usefulness of the filter in Eq. \eqref{eq:intro_defE} is that many physically distinct sources of nonadditivity become sharply distinguishable at the level of the large-$L$ scaling of ${\cal E}$. We develop this in three complementary settings.

\par The filter detects genuine nonextensivity whenever a nonintegrable static mediator couples to an operator with nonzero one-point function. Screening mechanisms in relativistic thermal QFT, such as the Debye mass or derivative couplings that kill the zero mode, typically prevent this. We use the field-theory formulation of an unscreened attractive mediator (a nonrelativistic scalar potential $\Phi$ coupled to mass density) to exhibit a controlled setting in which the sufficient IR criterion of the Nonextensivity criterion is satisfied and the standard thermodynamic limit fails (Sec.~\ref{sec:gravFT}). 

\par In gapped $(2+1)$-dimensional phases described at long distances by Chern--Simons (CS) topological effective field theory, the partition function on $\Sigma\times S^1$ contains a topology-dependent factor $\dim{\cal H}(\Sigma)$, independent of geometric moduli. In the low-temperature, large-size regime $L\gg\xi$, this yields
\begin{align}
	\Fcan(L,\beta)=&-\beta P(\beta)V-\log \dim{\cal H}(\Sigma) \nonumber \\
	 +&O(e^{-L/\xi})\ ,
\end{align}
and therefore the dimensionless replica energy $\beta{\cal E}$ converges to a universal constant,
\begin{align}
	\beta{\cal E}(L,\beta)=&-\log \dim{\cal H}(\Sigma)+O(e^{-L/\xi}),
\end{align}
as shown in Sec.~\ref{sec:topoE}. Thus $\beta{\cal E}$ isolates the purely topological $O(L^0)$ contribution to $\Fcan$, providing a thermodynamic diagnostic of topology-dependent nonadditivity that is invisible to bulk densities.

\par Fracton phases exhibit robust ground-state degeneracy that grows with linear system size rather than being $O(L^0)$.
For the X-cube model on a three-torus, $\log{\rm GSD}(L)\sim \gamma L$. In the same low-temperature, large-size regime one has
\begin{align}
	\Fcan(L,\beta)=-\beta P(\beta)L^3-\log{\rm GSD}(L)+\cdots,
\end{align}
and the replica-energy filter removes the bulk term but retains the subextensive contribution, giving
\begin{align}
	\beta{\cal E}(L,\beta)=&-\Big(1-\frac13L\partial_L\Big)\log{\rm GSD}(L)+\cdots \nonumber \\
	\sim & -\frac23\,\gamma L.
\end{align}
This identifies a distinct regime of subextensive additivity violation, where the thermodynamic limit exists ($\Fcan/L^3\to -\beta P$) but ${\cal E}$ diverges with $L$ (Sec.~\ref{sec:fracton}). Moreover, the full shape-sensitive identity in Eq. \eqref{eq:eulerdiff} naturally produces anisotropic replica energies ${\cal E}_i$ that diagnose additivity breaking along different directions.

\subsection{Relation to prior work and novelty}

Rigorous results on the existence of the thermodynamic limit for short-range systems (under stability and tempering hypotheses) are well known, and clarify why ordinary local QFTs recover extensivity at large volume in generic gapped situations \cite{ruelle}. On the other hand, long-range and nonadditive thermodynamics has a substantial literature, including formulations that introduce generalized Euler/Gibbs--Duhem relations and replica-energy-like quantities in nonadditive systems  \cite{Latella2015ThermodynamicsON, mori2015, Touchette_2002}. Our contribution is complementary: we provide a field-theoretic definition of replica energies directly in terms of $\Fcan=-\log Z$, derive an exact Ward/RG representation in Eq. \eqref{eq:intro_trace} for QFT, and show that the associated differential operator acts as a scaling filter that isolates universal subextensive data.

In $d=3$  \cite{FEletter} we refine this idea into a canonical double filter that removes all local power-law scheme ambiguities of $\Fcan_{S^3}=-\log |Z(S^3_R)|$ and defines a UV-finite function $F_{\cal E}(R)$. At conformal fixed points this quantity reduces to the sphere free energy ${\cal F}=-\log |Z(S^3)|$. Away from fixed points, however, it is not the same object as the entropic $F$-function of Ref.~\cite{CasiniHuerta:2012}. A central point of the present work is that the latter has a natural interpretation as a derivative of the replica-energy construction applied not to the bulk partition function, but to the replica-defect free energy
\begin{equation}
\Delta\Fcan_n(R)=\Fcan_n(R)-n\Fcan_1(R).
\end{equation}
Indeed, after the replica subtraction has removed the purely extensive bulk contribution, the first-order defect filter gives
\begin{equation}
F(R)=-\left(1-R\partial_R\right)\partial_n\Delta\Fcan_n(R)\big|_{n=1}.
\end{equation}
Thus the entropic $F$-function is minus the $n\to1$ derivative of the replica-energy filter applied to the entangling defect, whereas $F_{\cal E}(R)$ is the bulk sphere free energy passed through a second-order scaling filter. This distinction explains why the two quantities agree in their fixed-point data but need not have the same monotonicity properties along an RG flow.

In $d=4$ we apply the same filter idea to construct a UV-finite function ${\cal A}_{\cal E}(R)$ on the round four-sphere $S^4_R$, which equals $4a$ at RG fixed points and, for a free massive scalar field, is monotonically decreasing as a function of $mR$. We discuss the limiting values of ${\cal A}_{\cal E}(R)$ and compare the results to the 3d case.

We also use the double filter as a sharp discriminator between constant (liquid) topological order and linear subextensive fracton scaling. This framework cleanly distinguishes finite-size boundary effects, topology-dependent constants, and fracton subextensive divergences within a single thermodynamic/RG language.

\section{Extensivity and additivity}

\subsection{Definitions}
Consider a thermodynamic system $\mathcal{S}$ of volume $V$, particle number $N$, and inverse temperature $\beta=1/T$, where we set $k_B=1$. Its thermodynamic properties are encoded in the canonical partition function $Z=Z(V,N,\beta)$ and the dimensionless canonical free energy $\Fcan(V,N,\beta)=-\log Z=\beta F_{\rm can}$.

\begin{definition}[Extensivity]
A system is said to be extensive if $\Fcan$ satisfies the scaling law
\begin{equation}
\label{eq:ext0}
    \Fcan(\xi V, \xi N, \beta)=\xi \Fcan(V, N, \beta)
\end{equation}
for any $\xi > 0$. This rescaling must not change the shape of the volume $V$.
\end{definition}
\begin{definition}[Additivity]
    Let us divide a system $\mathcal{S}$ into two subsystems $\mathcal{A}$ and $\mathcal{B}$. The system $\mathcal{S}$ is said to be additive if
    \begin{equation}
        \Fcan(V_{\cal S}, N_{\cal S}, \beta)=\Fcan(V_{\cal A}, N_{\cal A}, \beta)+\Fcan(V_{\cal B}, N_{\cal B}, \beta)
    \end{equation}
    for any two subsystems ${\cal A}$ and ${\cal B}$.
\end{definition}
Extensivity implies that the system has a well-defined thermodynamic limit $V, N\to\infty$ for any shape of $V$. Additivity implies extensivity (since additivity holds for all subsystem decompositions, including those that preserve shape), and hence nonextensivity implies nonadditivity  \cite{Touchette_2002, mori2015}.

Consider again the system $\mathcal{S}$. The volume $V$ is $d$-dimensional and is characterized by the length scales $L_1,\dots, L_d$ normalized so that $V=L_1\dots L_d$. We write its dimensionless free energy as $\Fcan(\{L_i\},N,\beta)$. We define the generalized pressures  \cite{Mogliacci:2018oea}
\begin{equation}
    P_i=-\frac{1}{\beta}\frac{L_i}{V}\frac{\partial \Fcan}{\partial L_i}
\end{equation}
which measure the stress along each direction $i$ (i.e.\ the response of the free energy to a shape deformation
that stretches the system along $L_i$).
The generalized pressures are related to the ordinary pressure $P$ by
\begin{align}
\label{eq:p}
    P=-\frac{1}{\beta}\frac{\partial \Fcan}{\partial V}=\frac{1}{d}\sum_{i=1}^d P_i
\end{align}
Internal energy, chemical potential and the entropy are defined as usual
\begin{align}
\label{eq:umus}
    & U = \frac{\partial \Fcan}{\partial \beta}\ , &&\mu = \frac{1}{\beta}\frac{\partial \Fcan}{\partial N} \ , && S = \beta U-\Fcan
\end{align}
Let us perform on ${\cal S}$ the transformation
\begin{align}
    &N \mapsto \xi N \nonumber \\
    &L_i \mapsto \xi^{\frac{1}{d}+\alpha_i}L_i \quad \text{with}\quad \sum_{i=1}^d\alpha_i=0
\end{align}
The numbers $\alpha_i$ encode the change of the shape of the volume $V$. A system is extensive \textit{and} additive if
\begin{align}
\label{eq:scalingW}
    \Fcan(\{\xi^{\frac{1}{d}+\alpha_i}L_i\}, \xi N, \beta)=\xi \Fcan(\{L_i\}, N, \beta)
\end{align}
Differentiating with respect to $\xi$ around $\xi=1$ and using the definition of the thermodynamic quantities one obtains
\begin{equation}
\label{eq:geneuler}
    U-\frac{S}{\beta} +\left(P+\sum_{i=1}^d\alpha_iP_i\right)V-\mu N = 0
\end{equation}
Setting $\alpha_i=0$ for all $i=1,\dots, d$ one recovers the Euler relation, a hallmark of extensivity.

\subsection{Consequences and violations}

As an immediate consequence of extensivity combined with dimensional analysis, consider an extensive system with $d=3$, no particle-number term, and a dimensionless coupling $g$. Since $\Fcan$ is dimensionless, it can depend on its arguments only through $g$ and the ratio $\frac{V}{\beta^3}$. Extensivity implies that $\Fcan$ must be linear in the volume $V$. It follows that
\begin{align}
    \Fcan = f(g)\frac{V}{\beta^3}
\end{align}
that implies the Stefan-Boltzmann law $U\propto T^4$. Note that in this case the replica energy vanishes
identically, ${\cal E}=0$, consistent with exact extensivity at a conformal fixed point.
\par We now turn to the violations of additivity and extensivity. A nonadditive and/or nonextensive behaviour may be displayed by:
\begin{enumerate}[(a)]
    \item Short-range interacting systems with characteristic length comparable to the range of the interactions. In these systems, nonextensivity and/or nonadditivity disappear at large $V, N$.
    \item Long-range interacting systems. In these systems, nonextensivity and/or nonadditivity persist at large $V, N$.
\end{enumerate}
In nonadditive and/or nonextensive systems, the dimensionless canonical free energy does not satisfy the simple scaling law in Eq. \eqref{eq:scalingW}. The relation in Eq.\eqref{eq:geneuler} must be generalized to
\begin{align}
\label{eq:eulerviolation}
    U-\frac{S}{\beta} + \left(P+\sum_{i=1}^d\alpha_iP_i\right)V-\mu N = \mathcal{E}+\sum_{i=1}^d\alpha_i\mathcal{E}_i
\end{align}
where $\mathcal{E}$ and the $\mathcal{E}_i$ are called \textit{replica energies}. The \emph{isotropic} replica energy ${\cal E}$ quantifies the violation of extensivity, while \emph{anisotropic} replica energies ${\cal E}_i$ quantify the violations of additivity. Equation \eqref{eq:eulerviolation} can be recast as a differential relation involving $\Fcan$
{\small
\begin{align}
\label{eq:eulerdiff}
    \sum_{i=1}^d\left(\frac{1}{d}+\alpha_i\right)L_i\partial_{L_i}\Fcan+N\partial_N\Fcan=\Fcan-\beta\left({\cal E}+\sum_{i=1}^d\alpha_i{\cal E}_i\right)
\end{align}}
In the rest of this work, we set $\alpha_i=0$ for all $i=1,\dots,d$ and restrict to systems or thermodynamic states for which the $\mu N$ term is absent (for example, a theory without a conserved particle number, or a state with $\mu=0$). It is then convenient to express $\Fcan$ as a function of a single length scale $L$ such that $V=L^d$. The relations in Eqs.~\eqref{eq:eulerviolation} and \eqref{eq:eulerdiff} reduce to
\begin{align}
\label{eq:eulerspecial}
    & U-\frac{S}{\beta}+PV=\mathcal{E} \nonumber \\
    &\frac{1}{d}L\partial_{L}\Fcan=\Fcan-\beta{\cal E}
\end{align}
Indeed, the first line gives ${\cal E}=F_{\rm can}+PV$, while $\Fcan=\beta F_{\rm can}$ and $\beta PV=-\frac1dL\partial_L\Fcan$; the second line follows directly. This also fixes the overall sign of the replica-energy filter.
This relation encodes violations of extensivity when the particle-number term is absent. If the ratio $\frac{\beta{\cal E}}{V}$ does not vanish in the thermodynamic limit, there is a genuine violation of extensivity, and thus of additivity. On the contrary, for extensive systems in the large-volume limit one has
\begin{align}
\label{eq:extensive}
    &\Fcan(L, \beta) \underset{L\to\infty}{\sim} L^df(\beta) && \text{(extensive systems)}
\end{align}

\subsection{Filters}
\label{subsec:filter}
When the volume of the spatial manifold depends on the length scale $L$ alone, we can write the isotropic replica energy ${\cal E}$ as
\begin{align}
	\beta{\cal E} = \left(1-\tfrac{1}{d}L\partial_L\right)\Fcan
\end{align}
The differential operator ${\cal D}_L=\left(1-\frac{1}{d}L\partial_L\right)$ is called a \emph{filter}. It removes the purely extensive contributions to the canonical free energy:
\begin{align}
	{\cal D}_LL^d = 0
\end{align}
\par Consider a system whose long-distance physics is governed by an RG fixed point, perturbed by a set of scaling operators $\{{\cal O}_a\}$ with RG eigenvalues $y_a$ (so that under $L\to bL$ one has $g_a\to b^{y_a}g_a$). Standard finite-size scaling implies an expansion of the form \cite{Zinn-Justin:2002ecy, Collins:1984xc, Cardy:1996book}
\begin{align}
	\label{eq:fss_form}
	\Fcan(L,\beta;\{g_a\}) = -\beta P(\beta)\,L^d +\sum_a c_a(\beta)\,g_a\,L^{d+y_a} +\cdots,
\end{align}
where the dots include higher-order terms and possible logarithms associated with marginal directions. Applying $\mathcal{D}_L$ to Eq. \eqref{eq:fss_form} yields
\begin{align}
	\label{eq:filter_result}
	\beta{\cal E}(L,\beta) = \mathcal{D}_L \Fcan
	&= \sum_a \Big(1-\frac{d+y_a}{d}\Big) c_a(\beta)\,g_a\,L^{d+y_a} +\cdots\nonumber\\
	&= \sum_a -\frac{y_a}{d}\,c_a(\beta)\,g_a\,L^{d+y_a} +\cdots.
\end{align}
Therefore, the leading large-$L$ behavior of ${\cal E}$ is governed by the most slowly decaying (or growing) correction to extensivity. Equation \eqref{eq:filter_result} immediately yields a diagnostic classification:

\begin{enumerate}
	\item \textbf{Irrelevant perturbations} ($y_a<0$): $d+y_a<d$, so the correction in $\Fcan$ is $o(L^d)$ and
	$\beta{\cal E}/L^d\sim L^{y_a}\to 0$ (extensivity is restored at large $L$). In particular, a subextensive growth $\Fcan\sim L^s$ with $s<d$ corresponds to $y_a=s-d<0$ and is diagnosed by the filter, including the fracton regime of Sec. \ref{sec:fracton}.
	
	\item \textbf{Marginal directions} ($y_a=0$): the corresponding term may be $O(L^d\log L)$ or $O(\log L)$ depending on the observable and manifold so that $\mathcal{D}_L$ typically converts these into constant or logarithmic contributions. This includes the universal logarithms in $2$d critical systems  \cite{Cardy:1988tk, DiFrancesco:1997nk}.
	
	\item \textbf{Relevant perturbations} ($y_a>0$): the correction grows superextensively as $L^{d+y_a}$, and
	$\beta{\cal E}/L^d\sim L^{y_a}$ grows accordingly, signaling persistent scaling violations.
\end{enumerate}

In this sense, the replica energy is a ``finite-size RG filter'': it removes the dominant extensive piece and returns
directly the scaling content of the subleading corrections.

\subsection{Dilatation Ward identities}
\label{subsec:dward}

\subsubsection{Generalities}
Let us consider a quantum field theory for which the particle-number term is absent, on a $d$-dimensional spatial manifold $\Sigma_L$ with characteristic length scale $L$ and Euclidean flat metric $g_{\mu\nu}=\delta_{\mu\nu}$. We denote the elementary fields by $\phi$, dimensionless couplings by $g$, and squared masses by $m^2$. The renormalized thermal partition function of such a theory can be expressed in path-integral form
\begin{align}
\label{eq:pathint}
    Z\equiv \Tr[e^{-\beta H}] = \int [d\phi]\ e^{-S_E[\phi; g, m^2]}
\end{align}
where the trace Tr is taken over the Hilbert space of states. The Euclidean action functional $S_E$ can be expressed as the integral of the Euclidean lagrangian density
\begin{align}
    S_E[\phi; g, m^2]= \int\limits_{\Sigma_L\times S^1_{\beta}}d^{d+1}x\,\mathcal{L}_E[\phi; g, m^2]
\end{align}
where $S^1_{\beta}$ is the Euclidean-time circle of circumference (period) $\beta$, parametrized by $\tau\sim\tau+\beta$; if represented as a round circle, its geometric radius is $\beta/(2\pi)$. Inside the path integral \eqref{eq:pathint} the bosonic (fermionic) fields have periodic (antiperiodic) boundary conditions along $S^1_{\beta}$. Due to renormalization, the couplings and masses acquire a dependence on an energy scale $\mu$ that does not appear in the bare action. The dependence is encoded in two functions of the couplings, $\beta(g)$ and $\nu(g)$, defined by
\begin{align}
    &\beta(g)=-\mu\frac{dg}{d\mu},\nonumber\\
    &\nu(g)= 2-\mu\frac{d}{d\mu}\log m^2 .
\end{align}
Thus $\beta(g)$ and $\nu(g)$ are the length-flow beta function and the full
scaling exponent of $m^2$, respectively.
The functions $\beta(g)$ and $\nu(g)$ depend only on the short-distance structure of the theory and can be computed in a Euclidean flat $(d+1)$-dimensional space using dimensional regularization  \cite{tHooft:1972tcz}.
\par The dimensionless renormalized free energy $\Fcan=-\log Z$ depends on $\mu$ both explicitly and through the couplings \footnote{We are assuming that all IR divergences are regulated by $L$.}
\begin{align}
    \Fcan = \Fcan(L, \beta, g(\mu), m^2(\mu), \mu)
\end{align}
The response of $\Fcan$ under a rescaling of the manifold $\Sigma_L\times S^1_{\beta}$ is encoded in the dilatation Ward identity
\begin{align}
\label{eq:dilatation}
    \left(\beta\partial_{\beta}+L\partial_L\right)\Fcan=\int\limits_{\Sigma_L\times S^1_{\beta}} d^{d+1}x\ \langle{\Theta_{\mu}}^{\mu}\rangle
\end{align}
where $\Theta^{\mu\nu}$ is the improved stress-energy tensor  \cite{Callan:1970ze}. In flat space, its trace ${\Theta_{\mu}}^{\mu}$ satisfies the operator identity \footnote{The lhs of Eq. \eqref{eq:Theta} may contain also a four-divergence $\partial_{\mu}V^{\mu}$. In this paper, we will always assume that the boundary conditions of the fields and the topology of the manifold $\Sigma_L\times S_{\beta}^1$ are chosen so that this term vanishes when Eq. \eqref{eq:Theta} is inserted into Eq. \eqref{eq:dilatation}.}
\begin{align}
\label{eq:Theta}
    {\Theta_{\mu}}^{\mu} = \beta(g)\mathcal{O}_g+m^2\nu(g)\mathcal{O}_{m^2}
\end{align}
where $\mathcal{O}_g$ and $\mathcal{O}_{m^2}$ are the operators conjugate to the coupling $g$ and square mass $m^2$. Using Eqs. \eqref{eq:p}, \eqref{eq:umus} we express the resulting equation as
\begin{align}
\label{eq:virial}
    \beta \left(-U+d\cdot PV\right)=&-\beta(g)\int\limits_{\Sigma_L\times S^1_{\beta}}d^{d+1}x\, \expval{\mathcal{O}_g} \nonumber \\
    &-m^2\nu(g)\int\limits_{\Sigma_L\times S^1_{\beta}}d^{d+1}x\, \expval{\mathcal{O}_{m^2}}
\end{align}
This is nothing but the virial theorem for a quantum field-theoretical system: the left-hand side is the standard thermodynamic combination $-U+d\cdot PV$ that vanishes for a scale-free theory, while the right-hand side encodes the anomalous scaling contributions arising from UV renormalization.

\subsubsection{Replica energy from Ward identities and RG data}
\label{sec:EfromRG}

When the particle-number term is absent, the isotropic replica energy ${\cal E}$ is defined by Eq.~\eqref{eq:eulerspecial}:
\begin{align}
\label{eq:defE_simple}
\beta{\cal E}(L,\beta)=\Fcan(L,\beta)-\frac{1}{d}L\partial_L \Fcan(L,\beta).
\end{align}
We now derive an RG/trace representation of ${\cal E}$.

First, by definition of the operators conjugate to the renormalized couplings, differentiation of the path integral in Eq. \eqref{eq:pathint} yields the standard identities \cite{Collins:1984xc,Zinn-Justin:2002ecy}
\begin{align}
\label{eq:conjugate_ops}
&\partial_g \Fcan = \int_{\Sigma_L\times S^1_\beta} d^{d+1}x\,\langle {\cal O}_g\rangle,
\nonumber\\
&\partial_{m^2} \Fcan = \int_{\Sigma_L\times S^1_\beta} d^{d+1}x\,\langle {\cal O}_{m^2}\rangle.
\end{align}
Using the operator identity in Eq. \eqref{eq:Theta} for the trace of the improved stress tensor and inserting it into the dilatation Ward identity of Eq. \eqref{eq:dilatation}, one obtains the RG-type equation
\begin{align}
\label{eq:RGW}
\left(\beta\partial_\beta+L\partial_L\right)\Fcan
=
\beta(g)\,\partial_g \Fcan + m^2\nu(g)\,\partial_{m^2}\Fcan.
\end{align}
Eliminating $L\partial_L \Fcan$ between Eqs. \eqref{eq:RGW} and \eqref{eq:defE_simple} gives a compact expression for the replica
energy:
\begin{align}
\label{eq:E_master}
\beta{\cal E}
=
\Fcan+\frac{1}{d}\,\beta\partial_\beta \Fcan
-\frac{1}{d}\Big[\beta(g)\partial_g \Fcan+m^2\nu(g)\partial_{m^2}\Fcan\Big].
\end{align}
Using $\partial_\beta\Fcan=U$ and Eq. \eqref{eq:conjugate_ops}, Eq.~\eqref{eq:E_master} can be equivalently written as an
integrated trace relation,
\begin{align}
\label{eq:E_trace}
\beta{\cal E}
=
\Fcan+\frac{\beta}{d}U
-\frac{1}{d}\int_{\Sigma_L\times S^1_\beta}d^{d+1}x\, \langle{\Theta_\mu}^{\mu}\rangle.
\end{align}
Equation \eqref{eq:E_trace} makes explicit that ${\cal E}$ is a scaling filter: it vanishes for purely extensive
contributions to $\Fcan$ and isolates precisely the non-homogeneous pieces controlled by the trace Ward identity.

\subsection{Applications}

\subsubsection{Short-range interacting systems}
First, we recall a general fact: every quantum field theory at finite volume is both nonextensive and nonadditive. A nonadditive behavior is exhibited even by free field theories with Dirichlet  \cite{Mogliacci:2018oea} and periodic  \cite{Gliozzi:2007jh} boundary conditions. In the latter case, an explicit computation of the zero mode contribution to the partition function reveals that the replica energy scales as $\mathcal{E} \propto \log L $.
\par In interacting field theories additivity is broken by the propagators which depend nontrivially on $L$, see e.g. the scalar propagator in a periodic box with edges of length $L_1,\dots, L_d$
\begin{align}
    &\expval{\phi(x)\phi(y)}_{\rm free}= \sum_{\mathbf{n}\in \mathbb{Z}^{d+1}}\frac{1}{p_{\mathbf{n}}^2+m^2} e^{ip_{\mathbf{n}}\cdot (x-y)}
\end{align}
where the discrete momenta are 
\begin{align}
    p_{\mathbf{n}}=\left(\frac{2\pi n_0}{\beta},\frac{2\pi n_1}{L_1},\dots,\frac{2\pi n_d}{L_d}\right) 
\end{align}
Remarkably, in $(1+1)$-dimensional CFTs on a spatial circle the replica energy is controlled by the Casimir term and scales as $\beta{\cal E}\propto \beta/L$ (see Sec.~\ref{subsec:2d_CFT}). A $\log L$ dependence arises only in special cases such as free massless bosons with an unsuppressed zero mode, and is not a generic CFT feature.
\par However, in theories with short-range interactions additivity is restored at infinite volume, because in that limit propagators do not depend on the shape of $\Sigma_L$ and the system becomes translation-invariant. It follows that $\Fcan$, which is minus the sum of connected vacuum diagrams, depends on $L$ only through a proportionality factor $L^d$, as mentioned above.
\par Let us consider a translation-invariant system in the large-volume limit and express Eq. \eqref{eq:virial} in terms of the internal energy density $u=U/V$. Thanks to translation invariance the rhs simplifies
\begin{align}
\label{eq:virial_density}
    -u+d\cdot P=-\beta(g) \expval{\mathcal{O}_g} -m^2\nu(g)\expval{\mathcal{O}_{m^2}}
\end{align}
At least in principle, it is possible to have a residual, non-negligible $L$-dependence if the condensates are IR-divergent. In that case, at least one of $u$ and $P$ would be divergent as well, and hence $\Fcan$ would not satisfy Eq. \eqref{eq:extensive} even for $L\to\infty$. However, when $d>2$ and all elementary fields are massive we know \textit{a priori} that IR divergences are absent because all propagators are suppressed at large distances by exponential factors of the form $e^{-m|x-y|}$.

\subsubsection{Long-range interacting systems}
The prototypical example of long-range interacting model is QED, whose Euclidean Lagrangian density is
\begin{align}
    \mathcal{L}_E=\frac{1}{4}F_{\mu\nu}F^{\mu\nu}+\bar{\psi}(\slashed{D}+m)\psi
\end{align}
where the covariant derivative is
\begin{align}
    D_{\mu}\psi=\partial_{\mu}\psi-ieA_{\mu}\psi
\end{align}
In the Feynman gauge, Eq. \eqref{eq:virial_density} has the form
\begin{align}
-u+d\cdot P
=
+i\beta(e)\expval{\bar{\psi}\slashed{A}\psi}
-m\nu_m(e)\expval{\bar{\psi}\psi}.
\end{align}
For the fermion mass we analogously define $\nu_m(e)=1-\mu\,d\log m/d\mu$.
Eliminating the photon by means of its equations of motion, one finds
\begin{align}
\label{eq:virialqed}
    &-u+d\cdot P\nonumber \\
    =&-e\,\beta(e)\int\limits_{\Sigma_L\times S_{\beta}^1}d^{d+1}x\ \expval{\bar{\psi}\gamma^{\mu}\psi(x)\bar{\psi}\gamma^{\nu}\psi(0)}D_{\mu\nu}(x) \nonumber \\
    &-m\nu_m(e)\expval{\bar{\psi}\psi}
\end{align}
where $D_{\mu\nu}(x)$ is the \textit{full} photon propagator. Since the electron field $\psi$ is massive, we do not expect the condensate $\expval{\bar{\psi}\psi}$ to be IR-divergent. The story is different with the other term. First, we decompose the two-current correlator into its connected and disconnected part
\begin{align}
    \expval{\bar{\psi}\gamma^{\mu}\psi(x)\bar{\psi}\gamma^{\nu}\psi(0)}=&\expval{\bar{\psi}\gamma^{\mu}\psi(x)\bar{\psi}\gamma^{\nu}\psi(0)}_{\rm c} +\eta^2n^{\mu}n^{\nu}
\end{align}
where $\expval{\bar{\psi}\gamma^{\mu}\psi}=n^{\mu}\eta$. The connected part of the two-current correlator is exponentially suppressed by a factor $e^{-\pi |x|/\beta}$  due to the antiperiodic boundary conditions for fermions. The only possibly IR-divergent part of Eq. \eqref{eq:virialqed} is then
\begin{align}
\label{eq:irdiv}
    (-u+d\cdot P)\rvert_{\text{IR-div.}}=&-e\,\beta(e)\eta^2n^{\mu}n^{\nu}\int\limits_{\Sigma_L\times S_{\beta}^1}d^{d+1}xD_{\mu\nu}(x) \nonumber \\
    =&-e\,\beta(e)\eta^2n^{\mu}n^{\nu}\int\limits_{\Sigma_L}d^d\mathbf{x}\ D_{\mu\nu}^{(0)}(\mathbf{x})
\end{align}
where in the second line we evaluated the integral along $S_{\beta}^1$, whose effect was to isolate the zero mode $D_{\mu\nu}^{(0)}(\mathbf{x})$ of the photon propagator.

\subsubsection{A sufficient criterion for nonextensivity}
We are now in the position to state a sufficient criterion for the nonextensivity (and thus the nonadditivity) of a thermal QFT with the particle-number term absent.
\begin{remark*}
    Consider a thermal QFT with the particle-number term absent in $d$ spatial dimensions, consisting of a massless elementary field with full propagator $D(x)$ coupled linearly to a local operator $\mathcal{O}(x)$ that interpolates massive states. 
    Then, if
    \begin{align}
    \label{eq:condition}
        &\expval{\mathcal{O}}\neq 0,\nonumber\\
        &\int d^d\mathbf{x}\,\big|D^{(0)}(\mathbf{x})\big|=+\infty
    \end{align}
	where $D^{(0)}(\mathbf{x}) $ is the zero-mode of $D(x)$, and the induced static exchange is of definite sign (e.g.\ attractive) the theory is nonextensive, with the nonextensivity controlled by the integral
    \begin{align}
        \int d^d\mathbf{x}\ D^{(0)}(\mathbf{x})
    \end{align}
    A negative sign of this integral signals the lack of positivity of the Hilbert space, and hence a non-unitary dynamics.
\end{remark*}
The language in which this criterion is formulated is necessarily perturbative, because the distinction between "matter fields" and "force carriers" dissolves nonperturbatively. Thus, to apply this criterion to a theory we need to assume that the elementary fields in the action correspond to physical, weakly coupled degrees of freedom, maybe at the expense of renormalizability. This is not surprising, since nonextensivity and nonadditivity are controlled by the infrared structure of the theory.
\par Let us go back to QED: does it satisfy our Nonextensivity criterion? The answer is negative. At finite temperature, the only component of the electromagnetic current that condenses is $j^0$, because spatial rotation invariance forces $\langle\mathbf{j}\rangle=0$. Hence, nonextensivity depends only on the large-distance behaviour of $D_{00}^{(0)}(\mathbf{x})$. However, in finite-temperature QED, this two-point function is exponentially suppressed by the Debye mass  \cite{Kapusta:2006pm}. Thus, the condition in Eq. \eqref{eq:condition} is not satisfied.
\par Similar attempts to construct models that satisfy the Nonextensivity criterion seem to fail. A first candidate would be a relativistic gas of dipoles, where a neutral fermion is coupled to the photon via a term
\begin{equation}
    g \bar{\psi}\gamma^{\mu\nu}\psi F_{\mu\nu}
\end{equation}
In this model, however, the photon is coupled only to the \textit{spatial} components of the electromagnetic currents, that cannot develop a nonzero thermal expectation value due to rotational invariance. Similarly, we can try to couple a Goldstone boson $\pi$ to some matter current
\begin{align}
    g j^{\mu}\partial_{\mu}\pi 
\end{align}
The Goldstone boson is a promising force carrier, since the shift symmetry $\pi \mapsto \pi+\alpha $ protects it from the Debye screening even at finite temperature. What saves extensivity, is the fact that the derivative coupling kills the zero mode of $\pi$.
\par It seems that different phenomena always conspire to let relativistic thermal QFTs evade the Nonextensivity criterion presented above. This is not surprising, since the existence of the thermodynamic limit of statistical field theories has been proved rigorously under fairly general assumptions  \cite{ruelle}.
We summarize the evasion mechanisms in Table~\ref{tab:screening}.

\begin{table}[h]
	\begin{ruledtabular}
		\begin{tabular}{l l l}
		Model & Mediator & Evasion mechanism \\
		\hline
		QED & photon & Debye mass screens $D_{00}^{(0)}$ \\
		Dipole coupling & photon & $\langle \mathbf{j}\rangle=0$ by rotation invariance \\
		Goldstone/$\pi$ & $\pi$ & Derivative coupling kills zero mode \\
		\end{tabular}
	\end{ruledtabular}
	\caption{Mechanisms by which relativistic thermal QFTs evade the Nonextensivity criterion.}
	\label{tab:screening}
\end{table}

\subsubsection{A genuinely nonextensive field-theory model: unscreened attractive mediation}
\label{sec:gravFT}

We now exhibit a controlled counterexample in the language of field theory: a thermal system with a massless mediator coupled to a density operator with nonzero one-point function, and with an attractive sign such that screening does not restore extensivity.


Consider a nonrelativistic matter field $\psi$ in $d=3$ coupled to a scalar potential $\Phi$ (the Newtonian potential). The Euclidean action on $\Sigma_L\times S^1_\beta$ is
\begin{align}
	\label{eq:newton_action}
	S_E[\psi,\Phi]=\int_0^\beta d\tau\int_{\Sigma_L} d^3\mathbf{x}
	\Big[& \psi^\dagger(\partial_\tau-\frac{\nabla^2}{2m})\psi\nonumber\\
	&+\frac{1}{8\pi G}(\nabla\Phi)^2 -\Phi\,\rho(\psi) \Big],
\end{align}
with $\rho(\psi)\equiv m\,\psi^\dagger\psi$. The partition function is
\begin{align}
	Z=\int[d\psi\,d\psi^\dagger\,d\Phi]e^{-S_E[\psi,\Phi]}.
\end{align}
Functional-integral formulations of the self-gravitating gas of this type, and the associated thermodynamic subtleties, are discussed in classic reviews and detailed analyses (see e.g. Refs. \cite{Padmanabhan:1990, deVegaSanchez:2001, deVegaSanchez:2006}).

A key reason why the QED example fails the Nonextensivity criterion is Debye screening: finite-temperature polarization generates a positive static mass $m_D$ for the electric zero mode, so that the time-component of the photon propagator $D_{00}^{(0)}(\mathbf x)$ is exponentially suppressed. In the gravitational example the analogous mechanism is absent, for a simple sign reason: gravity has only one ``charge'' (mass), and the interaction is attractive, so linear response produces anti-screening rather than screening.

To see this in the same language used for Debye screening, consider the static ($\omega_n=0$) sector of the Euclidean theory \eqref{eq:newton_action} perturbatively expanded around a state with constant density $\bar{\rho}$. In momentum space the static propagator of the mediator becomes
\begin{align}
	\label{eq:Dphi0_rpa}
	D_{\Phi}^{(0)}(\mathbf k)= \frac{1}{\frac{\mathbf k^{\,2}}{4\pi G}-\Pi_{\rho\rho}(0,\mathbf k)}.
\end{align}
where 
$\Pi_{\rho\rho}(0,\mathbf k)$ is the static density--density correlator (compressibility kernel),
\begin{align}
	\Pi_{\rho\rho}(0,\mathbf k) = \int_0^\beta d\tau\int d^3\mathbf{x}\,e^{-i\mathbf k\cdot \mathbf x} \langle \delta\rho(\tau,\mathbf x) \delta\rho(0,0)\rangle_{\!c}\ \ge 0.
\end{align}
At small momentum $\Pi_{\rho\rho}(0,0) > 0$ for an ordinary compressible fluid. In QED the corresponding term enters with the opposite sign, producing a Yukawa denominator $\mathbf k^{\,2}+m_D^2$. Here, instead, the would-be ``screening mass'' is tachyonic:
\begin{align}
	\label{eq:mscr_tachyon}
	&m_{\rm scr}^2 \equiv -\,4\pi G\,\Pi_{\rho\rho}(0,0)\ <\ 0 \nonumber \\
	&D_{\Phi}^{(0)}(\mathbf k)\sim \frac{4\pi G}{\mathbf k^2-k_J^2}.
\end{align}
where $k_J = 2\pi/\ell_J$ is the Jeans wavenumber. The presence of this tachyonic pole signals the familiar gravitational instability of the homogeneous state  \cite{Padmanabhan:1990, deVegaSanchez:2001, deVegaSanchez:2006}. This interpretation is supported by the integration of the gravitational propagator over spacetime, which yields
\begin{align}
	\int d^3\mathbf{x}\ D^{(0)}_{\Phi}(\mathbf{x}) = D^{(0)}_{\Phi}(\mathbf{k}=0) = -\frac{4\pi G}{k_J^2}
\end{align}
The negativity of the rhs signals the lack of positivity of the Hilbert space \cite{Weinberg:1995mt}, and hence is a symptom of a nonunitary dynamics: thermal fluctuations destroy the homogeneous background around which our perturbative expansion is defined, due to its inherent instability.
\par We can control the divergence by putting it in a box of side $\sim \ell_J$. In this way the relevant integral for the Nonextensivity criterion becomes
\begin{align}
	\int_{\abs{\mathbf{x}}<\ell_J}d^3\mathbf{x}\ D^{(0)}_{\Phi}(\mathbf{x})\simeq \int_{\abs{\mathbf{x}}<\ell_J}d^3\mathbf{x}\int\frac{d^3\mathbf{k}}{(2\pi)^3} \frac{4\pi G}{\mathbf{k}^2-k_J^2} e^{i\mathbf{k}\mathbf{x}}
\end{align}
Let us compute this integral. We can expand the phase factor as a sum of spherical Bessel functions
\begin{align}
	e^{i\mathbf{k}\mathbf{x}} = \sum_{\ell = 0}^{\infty}(2\ell+1)i^\ell j_\ell (kr)
\end{align}
Since the integral is spherically symmetric we can retain only the first term
\begin{align}
	e^{i\mathbf{k}\mathbf{x}} = j_0(kr)+\text{non-spherically symmetric terms}
\end{align}
where
\begin{align}
	j_0(kr) = \frac{\sin (kr)}{kr}
\end{align}
Integrating over the coordinates we obtain
\begin{align}
	&\int_{\abs{\mathbf{x}}<\ell_J}d^3\mathbf{x}\ D^{(0)}_{\Phi}(\mathbf{x}) \nonumber \\
	=&\frac{2}{\pi}\int_0^\infty dk\frac{1}{k(k^2-k_J^2)}\left(\sin \frac{2\pi k}{k_J}+\frac{2\pi k}{k_J}\cos \frac{2\pi k}{k_J}\right)
\end{align}
We see that this integral is logarithmically divergent due to the simple pole of the integral near $k=k_J$.

\section{Topological replica energy in $(2+1)$ dimensions}
\label{sec:topoE}

The analysis so far has focused on (non)extensivity driven by infrared physics of propagating modes, and in particular on the sufficient criterion stated in the Nonextensivity criterion, which requires a non-integrable static propagator coupled to an operator with nonzero thermal one-point function. In this section we highlight a different and, in $(2+1)$ dimensions, very common source of finite-volume violations of Euler homogeneity: a topological sector. The resulting effect is subextensive (hence compatible with the existence of the thermodynamic limit in the sense of Ruelle  \cite{ruelle}), but it is robust and universal, and it is cleanly isolated by the replica energy ${\cal E}$ defined in Eq. \eqref{eq:eulerspecial}.

\subsection{Gapped phases, finite size, and topological sectors}

We work in $d=2$ spatial dimensions, with the particle-number term absent and $\alpha_i=0$, so that the replica energy is defined by Eq. \eqref{eq:eulerspecial}:
\begin{align}
	\label{eq:EdefTopoPed}
		\beta{\cal E}(L,\beta)\equiv \Fcan(L,\beta)-\frac{1}{2}\,L\partial_L \Fcan(L,\beta)
\end{align}
We take the theory to be in a gapped phase with correlation length $\xi<\infty$. Consider a spatial manifold $\Sigma_L$ of characteristic linear size $L$ and volume $V\propto L^2$, and assume
\begin{align}
	\label{eq:scalesTopoPed}
	L\gg\xi
\end{align}
Under these conditions the partition function is dominated by the low-energy sector, with exponentially small corrections from both thermal excitations and finite-size effects. In an ordinary topologically trivial gapped phase,
	the low-energy sector consists of a unique ground state on a closed $\Sigma$ and the leading contribution to $\Fcan$ is extensive:
\begin{align}
	\Fcan(L,\beta)=-\beta P(\beta)\,V + o(V),
\end{align}
so that ${\cal E}/V\to 0$ trivially.

In a topologically ordered gapped phase, the low-energy sector depends on the topology of $\Sigma$ and may be
degenerate even at infinite $L$. A canonical effective description of such phases in $(2+1)$ dimensions is provided by Chern--Simons topological field theory (possibly coupled to massive matter). The key point for our purposes is that the Euclidean partition function on $\Sigma\times S^1$ contains a purely topological factor given by the dimension of the topological Hilbert space on $\Sigma$:
\begin{align}
	\label{eq:ZtopoDimPed}
	Z_{\rm top}(\Sigma\times S^1)=\dim{\cal H}(\Sigma),
\end{align}
a standard result of CS/TQFT quantization\footnote{the path integral on $\Sigma\times S^1$ computes the trace of the identity on ${\cal H}(\Sigma)$ \cite{Witten:1989jones, Elitzur:1989csw}}.

\subsection{Factorization at low temperature and large size}

The scale separation in Eq. \eqref{eq:scalesTopoPed} implies that the full partition function admits the asymptotic factorization
\begin{align}
	\label{eq:ZfactTopoPed}
	Z(\Sigma_L\times S^1_\beta)
	=
	&\exp \big[\beta P(\beta) V\big]\dim{\cal H}(\Sigma)\Big[1+O(e^{-L/\xi})\Big].
\end{align}
The first factor is the standard extensive contribution controlled by the pressure while the second factor is topological and independent of geometric moduli such as $L$ (as long as the bulk remains gapped) and the bracketed terms collect subleading corrections.

	Taking minus the logarithm gives the corresponding decomposition of $\Fcan$:
\begin{align}
\label{eq:WdecompTopoPed}
	\Fcan(L,\beta) = -\beta P(\beta) V-\log \dim{\cal H}(\Sigma)+O(e^{-L/\xi}).
\end{align}

For Abelian $K$-matrix CS theories relevant to fractional quantum Hall states, the ground-state degeneracy on the torus $\Sigma=T^2$ is $|\det K|$ and in particular, for $U(1)_k$ CS one finds $\dim{\cal H}(T^2)=k$. A pedagogical derivation and further examples can be found in Ref. \cite{Tong:2016qhe}.

A subtlety in the microscopic derivation of $\dim{\cal H}(\Sigma)$ arises from the parity anomaly: integrating out massive fermions generates parity-odd terms in the effective action whose perturbative expression can be temperature-dependent, while large gauge invariance constrains the consistent description \cite{Deser:1997parity, DasDunne:1998largegauge}. In the low-temperature regime of Eq. \eqref{eq:scalesTopoPed}, such effects renormalize the effective topological data (e.g.\ level shifts) entering $\dim{\cal H}(\Sigma)$ and hence feed directly into the replica energy in Eq. \eqref{eq:EtopoPedMain}.

\subsection{Replica energy as a projector onto topology-dependent terms}

	We now apply the definition of Eq. \eqref{eq:EdefTopoPed} to the asymptotic form in Eq. \eqref{eq:WdecompTopoPed}. Since $V\propto L^2$,
\begin{align}
	\label{eq:LdLV}
		\frac{1}{2}\,L\partial_L\big(-\beta P(\beta)V\big)=-\beta P(\beta)V,
\end{align}
	and therefore the extensive contribution cancels identically inside Eq. \eqref{eq:EdefTopoPed}. The topological term $-\log\dim{\cal H}(\Sigma)$ is $L$-independent and survives unchanged. One finds
\begin{align}
	\label{eq:EtopoPedMain}
	\beta{\cal E}(L,\beta)
	&=
		-\log \dim{\cal H}(\Sigma) +O(e^{-L/\xi}),\nonumber\\
	&\Rightarrow \frac{{\cal E}(L,\beta)} {V} \xrightarrow[L\to\infty]{}0.
\end{align}
Eq. \eqref{eq:EtopoPedMain} is the main result of this section. It shows that ${\cal E}$ isolates the universal
	topological contribution to $\Fcan$ in a gapped $(2+1)$-dimensional phase: the replica energy is subextensive but nonvanishing, and it depends only on the topology of $\Sigma$ through $\dim{\cal H}(\Sigma)$.

	The existence of an $O(L^0)$, topology-dependent term in $\Fcan$ implies a robust finite-volume violation of strict additivity:
	there is no way to assign to two macroscopic subsystems a decomposition of $\Fcan$ that is independent of global topology and gluing data. In other words, even when the bulk is local and gapped (hence extensive in the thermodynamic limit), the partition function retains a ``global'' contribution that cannot be captured by an integral of local densities. This is precisely the kind of finite-volume nonadditivity that the replica-energy operator in Eq. \eqref{eq:EdefTopoPed} is designed to diagnose.

\subsection{Relation to conformal edge physics and logarithmic terms}

	If $\Sigma$ has a boundary or if the low-energy theory supports protected gapless edge modes, the low-energy sector generically contains a $(1+1)$-dimensional CFT on the boundary. In that case $\Fcan$ may receive additional universal subextensive contributions, including logarithmic finite-size terms of the kind discussed in two-dimensional critical systems \cite{Cardy:1988tk, DiFrancesco:1997nk}. The replica energy operator in Eq. \eqref{eq:EdefTopoPed} continues to project out the leading extensive piece, leaving the universal boundary/topological contributions as the dominant signal at large $L$.

We stress that the mechanism isolated here is not the long-range nonextensivity mechanism of the Nonextensivity criterion. In Maxwell--Chern--Simons(-matter) plasmas, the CS term gaps the gauge sector, and finite-temperature polarization effects typically render static correlators integrable at long distances. A systematic discussion of screening in Abelian CS theories at finite temperature can be found in Ref. \cite{AlvesDasPerez:2002screening}. Thus CS topological phases generically realize a robust subextensive replica energy, rather than a violation with
$\beta{\cal E}/V\not\to 0$.

\section{Subextensive nonadditivity in fracton phases}
\label{sec:fracton}

The previous sections distinguished two broad mechanisms for violations of additivity/extensivity:
finite-size effects in short-range systems, which disappear as $L\to\infty$ and
 infrared mechanisms in genuinely long-range systems, which may survive in the thermodynamic limit.
Here we emphasize a third, intermediate regime, realized by fracton phases and more generally by gapped phases
with subextensive ground-state degeneracy. In such systems the thermodynamic limit exists in the usual sense
	(i.e.\ $\Fcan/V$ converges), yet the replica energy ${\cal E}$ defined in Eq.~\eqref{eq:eulerspecial} diverges with $L$
in a controlled way.

\subsection{Low-temperature form of $\Fcan$ and subextensive degeneracy}
\label{subsec:liquid}

	We work with the particle-number term absent, in $d=3$ spatial dimensions, and consider a gapped local Hamiltonian system on a three-torus of linear size $L$ (a cubic box with periodic boundary conditions). Let $\xi$ be the correlation length. In the regime
\begin{align}
	\label{eq:fracton_regime}
	L, \beta\gg \xi,
\end{align}
thermal contributions from excited states are suppressed by $e^{-\beta/\xi}$ and finite-size corrections from bulk
correlations are suppressed by $e^{-L/\xi}$. Then the partition function factorizes as  \cite{Xfracton, Vijay:2016fracton}
\begin{align}
	\label{eq:fracton_fact}
	Z(L,\beta)=\exp\!\big[\beta P(\beta)\,L^3\big]
	{\rm GSD}(L)\Big[1+O(e^{-L/\xi})\Big],
\end{align}
where ${\rm GSD}(L)$ is the (robust) ground-state degeneracy on the chosen manifold.

For conventional gapped phases and for liquid topological phases, ${\rm GSD}(L)\sim O(L^0)$ (topology-dependent but
size-independent). By contrast, a hallmark of type-I (``foliated'') fracton phases is that ${\rm GSD}(L)$ grows
exponentially with system size, so that $\log {\rm GSD}(L)$ is subextensive but divergent as $L\to\infty$.
A canonical example is the X-cube model, whose degeneracy on a $L_x\times L_y\times L_z$ three-torus satisfies
\begin{align}
	\label{eq:xcube_gsd_aniso}
	\log_2{\rm GSD}=2L_x+2L_y+2L_z-3,
\end{align}
so that for $L_x=L_y=L_z=L$ one has $\log {\rm GSD}(L)=\gamma L+O(L^0)$ with $\gamma=6\log 2$  \cite{Vijay:2016fracton, Slagle:2017xcubeQFT, Seiberg:2020wsg}.

	Taking minus the logarithm of Eq. \eqref{eq:fracton_fact} yields
\begin{align}
	\label{eq:W_fracton_scaling}
	\Fcan(L,\beta)= -\beta P(\beta)\,L^3-\log{\rm GSD}(L)+O(e^{-L/\xi}).
\end{align}

\subsection{Replica energy and linear scaling for foliated fracton order}

	For $d=3$ spatial dimensions with the particle-number term absent, Eq.~\eqref{eq:eulerspecial} implies
\begin{align}
	\label{eq:fracton_E_operator}
		\beta{\cal E}(L,\beta)=\Big(1-\frac13 L\partial_L\Big)\Fcan(L,\beta).
\end{align}
Applying Eq. \eqref{eq:fracton_E_operator} to Eq. \eqref{eq:W_fracton_scaling} annihilates the extensive bulk term exactly:
\begin{align}
		\Big(1-\frac13L\partial_L\Big)\big[-\beta P(\beta)L^3\big]=0.
\end{align}
Hence the leading behavior of ${\cal E}$ is controlled by the subextensive piece $\log{\rm GSD}(L)$:
\begin{align}
	\label{eq:E_fracton_general}
		\beta{\cal E}(L,\beta)=&-\Big(1-\frac13 L\partial_L\Big)\log{\rm GSD}(L)+O(e^{-L/\xi}).
\end{align}
In particular, for a foliated fracton phase with
\begin{align}
	\label{eq:fracton_linear}
	\log{\rm GSD}(L)=\gamma L+O(L^0),
\end{align}
one obtains the asymptotic result
\begin{align}
	\label{eq:E_fracton_linear}
		\beta{\cal E}(L,\beta)=&-\frac23\,\gamma L+O(L^0)+O(e^{-L/\xi}),
	\nonumber\\
	&\Rightarrow
	\frac{{\cal E}(L,\beta)}{L^3}\xrightarrow[L\to\infty]{}0.
\end{align}
Equation \eqref{eq:E_fracton_linear} is the precise sense in which fracton phases realize subextensive nonadditivity: the thermodynamic limit is extensive, yet the replica energy diverges with $L$.

The size dependence of ${\rm GSD}(L)$ is not a surface correction in the conventional sense but rather it reflects
additional nonlocal constraints associated with subsystem symmetries/foliation structure. This is made explicit by the foliation-based characterization of type-I fracton order and by the renormalization-group viewpoint in which changing system size involves adding/removing layers of 2D topological resource states  \cite{Shirley:2018manifolds, Shirley:2019gauging}. From a continuum field-theory perspective, the same phenomenon appears as a sensitivity to the ultraviolet cutoff in the ground-state sector (a controlled ``taming of the infinity''), which is another manifestation of the UV/IR interplay in these phases  \cite{Slagle:2017xcubeQFT}.

\subsection{Anisotropic replica energies and the X-cube model}
\label{sec:fracton_aniso}

Fracton degeneracy is naturally anisotropic, so it is instructive to use the full shape-sensitive identity in Eq.
\eqref{eq:eulerdiff} rather than the isotropic specialization. Consider $d=3$ with the particle-number term absent and a rectangular three-torus with linear sizes $(L_x,L_y,L_z)$. In the low-temperature, large-size regime of Eq. \eqref{eq:fracton_regime}, the dimensionless canonical free energy takes the form
\begin{align}
	\label{eq:W_aniso_ansatz}
	\Fcan(L_x,L_y,L_z;\beta)=&-\beta P(\beta)\,V-\kappa_x L_x-\kappa_y L_y-\kappa_z L_z-\kappa_0\nonumber\\
	&+O(e^{-L/\xi}),
\end{align}
with $V=L_xL_yL_z$ and coefficients $\kappa_i$ determined by $\log{\rm GSD}$. For the X-cube model on the three-torus one has \cite{Vijay:2016fracton}
\begin{align}
	\label{eq:xcube_logGSD}
	\log {\rm GSD} = (2L_x+2L_y+2L_z-3)\log 2,
\end{align}
so that $\kappa_x=\kappa_y=\kappa_z=2\log 2$ and $\kappa_0=-3\log 2$.

We now insert Eq. \eqref{eq:W_aniso_ansatz} into the general differential identity of Eq.~\eqref{eq:eulerdiff} (with the particle-number term absent and $\sum_i\alpha_i=0$). Using $L_i\partial_{L_i}V=V$ and $\sum_i\alpha_i=0$, the $\alpha_i$-dependent part of Eq. \eqref{eq:eulerdiff} isolates the traceless combinations of the additivity-violating replica energies ${\cal E}_i$. Choosing the convention with no common additive shift gives
\begin{align}
	\label{eq:Ei_from_linear}
	&{\cal E}_x= \frac{\kappa_x}{\beta}L_x \nonumber \\
	&{\cal E}_y= \frac{\kappa_y}{\beta}L_y \nonumber \\
	&{\cal E}_z= \frac{\kappa_z}{\beta}L_z,
\end{align}
up to exponentially small corrections. The isotropic replica energy ${\cal E}$ then follows from the $\alpha_i=0$ projection of Eq. \eqref{eq:eulerdiff}:
\begin{align}
	\label{eq:E_from_linear}
	\beta{\cal E} = &-\kappa_0-\frac{2}{3}\left(\kappa_xL_x+\kappa_yL_y+\kappa_zL_z\right)+O(e^{-L/\xi}).
\end{align}
For the X-cube model in Eq. \eqref{eq:xcube_logGSD}, this yields
\begin{align}
	\label{eq:E_xcube_final}
	\beta{\cal E} = 3\log 2-\frac{4\log 2}{3}\left(L_x+L_y+L_z\right)+O(e^{-L/\xi}),
\end{align}
and in the cubic case $L_x=L_y=L_z=L$ one recovers $\beta{\cal E}\sim -4(\log 2)\,L$.

Equations \eqref{eq:Ei_from_linear}--\eqref{eq:E_xcube_final} show that fracton phases exhibit a controlled, universal
subextensive violation of additivity: the thermodynamic limit remains extensive ($\Fcan/V\to -\beta P$), yet the
replica energies ${\cal E}_i$ and ${\cal E}$ diverge with system size, reflecting the foliation/subsystem structure.

\subsection{General scaling relation}

The preceding computation relies only on scaling. Let $d$ be the spatial dimension and assume that at large $L$,
\begin{align}
	\label{eq:lemma_ansatz}
	\Fcan(L,\beta)=-\beta P(\beta)L^d + A(\beta)\,L^{s}+o(L^{s}),
	\qquad 0\le s<d.
\end{align}
Then, from $\beta{\cal E}=(1-\frac1dL\partial_L)\Fcan$, one finds
\begin{align}
	\label{eq:lemma_result}
	\beta{\cal E}(L,\beta) = \Big(1-\frac{s}{d}\Big)A(\beta)\,L^{s}+o(L^{s}).
\end{align}
Thus $s=0$ reproduces the topological regime with constant replica energy, while $s=1$ in $d=3$ reproduces the fracton regime of Eq.~\eqref{eq:E_fracton_linear}. For a degeneracy contribution, $A<0$, so the corresponding replica energy is negative. The exponent $s$ can be interpreted as the effective dimension of the manifold on which the nonadditive contribution is extensive.


\section{Replica energy at RG fixed points}
\label{sec:cF}

In this section we sharpen the connection between the replica-energy filter
\begin{align}
	\mathcal{D}_L \equiv \Big(1-\tfrac{1}{d}L\partial_L\Big), \qquad 	\beta{\cal E}=\mathcal{D}_L \Fcan,
\end{align}
and standard measures of degrees of freedom along RG flows.
The key point is that at conformal fixed points the leading finite-size terms in $\Fcan=-\log Z$ are fixed by symmetry, and $\mathcal{D}_L$ isolates precisely their universal coefficients.

\subsection{Two-dimensional CFTs: extracting the central charge}

\label{subsec:2d_CFT}
Consider a $(1+1)$-dimensional relativistic QFT at an RG fixed point (a 2D CFT) placed on a spatial circle
of circumference $L$ and Euclidean time circle of length $\beta$. In the low-temperature regime $\beta\gg L$, the partition function is dominated by the Casimir
energy on the circle, and one has the universal finite-size correction
\begin{align}
	E_0(L)=e_\infty L - \frac{\pi c}{6\,L} + o(L^{-1}),
\end{align}
hence
\begin{align}
	\Fcan(L,\beta)= \beta E_0(L) = \beta e_\infty L - \frac{\pi c}{6} \frac{\beta}{L}+o(\beta/L).
\end{align}
The $O(L)$ term is extensive (and scheme-dependent through $e_\infty$), while the $c$-term is universal. Applying the replica-energy filter at $d=1$,
\begin{align}
	\beta{\cal E}(L,\beta)
	=\Big(1-L\partial_L\Big)\Fcan(L,\beta)
	=-\frac{\pi c}{3}\frac{\beta}{L}+o(\beta/L),
\end{align}
so that the central charge is extracted by
\begin{align}
	\label{eq:c_extract}
	c=-\lim_{\beta/L\to\infty}\frac{3L}{\pi\beta}\,\beta{\cal E}(L,\beta).
\end{align}

While Eq.~\eqref{eq:c_extract} is an endpoint extraction at a fixed point, in two spacetime dimensions one can relate the flow of the dimensionless combination
\begin{align}
	c_{\cal E}(L)\equiv -\lim_{\beta/L\to\infty}\frac{3L}{\pi\beta}\,\beta{\cal E}(L,\beta)
\end{align}
to the universal spectral representation of stress-tensor correlators. In particular, unitarity (reflection positivity) implies the existence of a non-negative spectral density for the stress-tensor two-point function, and the net change of central charge along any
unitary RG flow obeys the dispersive sum rule\footnote{This is the spectral form of the
Zamolodchikov--Cardy sum rule.}
\begin{align}
	\label{eq:c_sumrule_spectral}
	\Delta c &\equiv c_{\rm UV}-c_{\rm IR}
	=\int_0^\infty\! d\mu\, c_I(\mu)\nonumber\\
	&=3\pi\int_{\mathbb{R}^2}\! d^2x\,|x|^2\,\langle \Theta(x)\Theta(0)\rangle,
	\qquad c_I(\mu)\ge0.
\end{align}
Thus $\Delta c>0$ for any nontrivial unitary flow, and $c_{\cal E}(L)$ provides a natural finite-size realization of a running ``degree-of-freedom counter'' whose endpoints reproduce $c_{\rm UV}$ and $c_{\rm IR}$  \cite{CappelliFriedanLatorre:1991}.

The underlying finite-size formula is classic in 2D conformal finite-size scaling and the conformal anomaly,
see Refs. \cite{BloteCardyNightingale:1986, Affleck:1986, Cardy:1996book}. Monotonicity is then inherited at the level of fixed points from Zamolodchikov's $c$-theorem: for any unitary RG flow between 2D fixed points,
\begin{align}
	c_{\rm UV}\ge c_{\rm IR},
\end{align}
and therefore the fixed-point value of the quantity in Eq. \eqref{eq:c_extract} decreases along the flow  \cite{Zamolodchikov:1986}.

\subsection{Flows on $S^3$: a universal thermodynamic $F$-function from a double filter}
\label{subsec:Ftheorem_precise}

\subsubsection{Construction and (lack of) monotonicity}
\begin{figure}[h]
    \centering
    \includegraphics[width=\columnwidth]{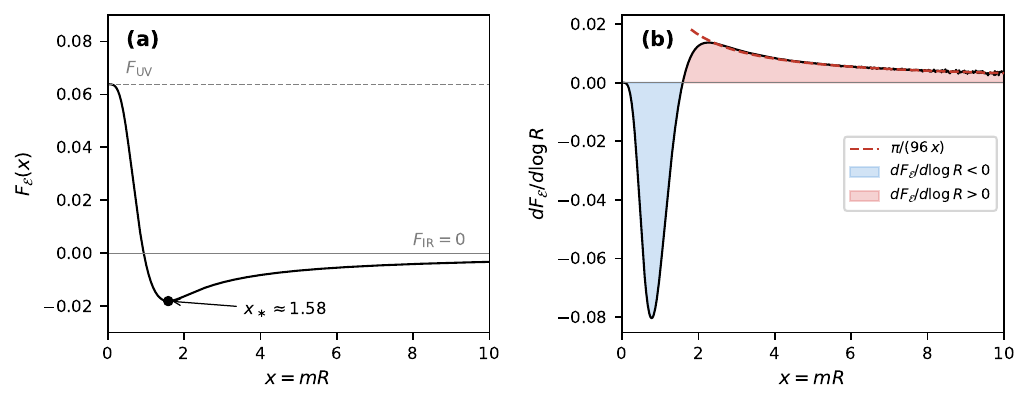}
    \caption{
    (a) Thermodynamic $F$-function $F_{\cal E}(x)$ for a free massive scalar on $S^3$ ($x=mR$).
    $F_{\cal E}$ starts at $F_{\rm UV}\approx 0.064$ (dashed line), overshoots below
    $F_{\rm IR}=0$, reaching a minimum at $x_\ast\approx 1.58$ (dot), then returns to $0^-$ from below. (b) Flow derivative $dF_{\cal E}/d\log R$.
    Negative for $x<x_\ast$, positive for $x>x_\ast$ (shaded regions). The dashed curve on the top right corner is the analytic asymptotic $\pi/(96\,x)$ in Eq.~\eqref{eq:large_x}, which controls the large-$x$ tail. The sign change demonstrates that $F_{\cal E}$ is not a monotone $F$-function.
    \label{fig:FE}}
\end{figure}

On $S^3_R$ the only UV-sensitive local counterterms allowed by diffeomorphism invariance are~\cite{BirrellDavies:1982,Vassilevich:2003}
\begin{align}
\Fcan_{\rm loc}=\lambda_0\!\int\!\sqrt{g}+\lambda_1\!\int\!\sqrt{g}\,\mathcal{R}\,,
\end{align}
 these two counterterms scale as $R^3$ and $R$ respectively. We define the differential operator $D\equiv R\,\partial_R$ and the double filter
\begin{align}
    \label{eq:double_filter}
    \mathcal{D}_{\rm ren}^{(3)}\equiv
    \Big(1-D\Big)\!\Big(1-\tfrac{1}{3}D\Big)
    =1-\tfrac{4}{3}D+\tfrac{1}{3}D^2,
\end{align}
which is the unique polynomial of degree $\le 2$ in $D$ satisfying $\mathcal{D}_{\rm ren}^{(3)}[R^3]=\mathcal{D}_{\rm ren}^{(3)}[R]=0$ and $\mathcal{D}_{\rm ren}^{(3)}[1]=1$. The thermodynamic $F$-function is
\begin{align}
    \label{eq:FE_def}
    F_{\cal E}(R)\equiv \mathcal{D}_{\rm ren}^{(3)}\,\Fcan_{S^3}(R).
\end{align}
By construction, $F_{\cal E}$  reduces to $F=-\log|Z(S^3)|$ at conformal fixed points. In Ref. \cite{FEletter} it has been shown that $F_{\cal E}$ decreases near the UV at leading nontrivial order in conformal perturbation theory. Consider a CFT on $S^3_R$ deformed by a relevant scalar $\mathcal{O}$ of dimension $\Delta<3$, with coupling $g$ of mass dimension $3-\Delta$. Let $\langle\mathcal{O}(x)\mathcal{O}(y)\rangle=C_{\mathcal{OO}}\big/\big(2R\sin\tfrac{\theta}{2}\big)^{2\Delta}$ be the conformal $2$ pt. function of this scalar, where $\theta$ is the geodesic angle between $x$ and $y$.  A perturbative calculation at second order on $g$ shows that
\begin{align}\label{eq:deriv}
    \frac{dF_{\cal E}}{d\log R}\Bigg\rvert_{O(g^2)}=-\kappa(\Delta)\,\frac{1}{2}g^2 C_{\mathcal{OO}}\,R^{6-2\Delta}<0
\end{align}
with
\begin{align}
    \kappa(\Delta) = \frac{64}{3}\pi^{7/2}2^{-2\Delta}\frac{\Gamma\left(\tfrac72-\Delta\right)}{\Gamma(3-\Delta)} > 0
\end{align}
This calculation probes the theory only next to a conformal fixed point, i.e. when $x\equiv mR \ll 1$. Away from fixed points an exact calculation of $\Fcan=\frac{1}{2}\Tr\log (-\nabla^2+m^2)$ is required. In \cite{FEletter} this calculation was performed with an appropriate diffeomorphism-preserving regularization, and it led to
\begin{align}
    D\Fcan
    =-\frac{\pi x^2}{2}\nu(x)\coth\big(\pi\nu(x)\big),
    \quad
    \nu(x)\equiv\sqrt{x^2-\tfrac14}.
\end{align}
At large $x$, $D\Fcan\sim-\frac\pi2 x^3+\frac{\pi}{16}x +\frac{\pi}{256}x^{-1} +\cdots$, so the filter kills $x^3$ and $x$. Therefore, since $\mathcal{D}_{\rm ren}^{(3)}[x^{-1}]=\tfrac{8}{3}\,x^{-1}$, the $\tfrac{\pi}{256}x^{-1}$ term gives the leading survivor:
\begin{align}
    \label{eq:large_x}
    \frac{dF_{\cal E}}{d\log R}\xrightarrow{x\to\infty}
    \frac{\pi}{96\,x}+O(x^{-3})>0.
\end{align}
Since the derivative is negative in the small $x$ regime by the perturbative analysis, as shown in Eq.~\eqref{eq:deriv}, and positive for large $x$ as in Eq.~\eqref{eq:large_x}, it must change sign. A numerical evaluation (Fig.~\ref{fig:FE}) gives the zero crossing at $x_\ast\equiv mR_\ast\approx 1.584$. Integrating from $x=0$ where $F_{\cal E}=F_{\rm UV}\approx 0.0638$, the function reaches
\begin{align}
    F_{\cal E}(x_\ast)\approx -0.0181 \approx -0.28\,F_{\rm UV},
\end{align}
overshooting below $F_{\rm IR}=0$ before slowly returning as $F_{\cal E}(x)\to -\pi/(96\,x)\to 0^-$. 

\subsubsection{Relation to the entropic $F$-function}
We now relate our results to the monotonicity proof of the $3$-sphere entanglement entropy computed in \cite{CasiniHuerta:2012} with the aid of the replica trick \cite{Prihadi:2022lsw, Callan:1994py, Holzhey:1994we, Calabrese:2004eu, Nishioka:2009un}. In presenting the latter, we will follow the presentation of Ref. \cite{Nishioka:2009un}. Consider a field theory with a set of elementary fields that we collectively denote as $\phi(x) = \phi(\mathbf{x}, t)$ living on a manifold $\mathcal{M}\times \mathbb{R}_t $ with $\mathcal{M}$ being $d$-dimensional. We consider a subset $A \subset \mathcal{M}$ and the reduced density matrix
\begin{align}
    [\rho_A]_{\phi_+\phi_-} = &\frac{1}{Z_1}\int  [d\phi] e^{-S_E[\phi]} \nonumber \\
    &\prod_{\mathbf{x}\in A}\delta(\phi(\mathbf{x},0^+)-\phi_+(\mathbf{x}))\delta(\phi(\mathbf{x},0^-)-\phi_-(\mathbf{x}))
\end{align}
with $Z_1$ chosen so that $\Tr \rho_A = 1 $. The entanglement entropy $S_A$ is defined as
\begin{align}
    S_A = -\Tr \rho_A \log \rho_A = -\frac{\partial}{\partial n}\Tr \rho_A^n\bigg\rvert_{n = 1}
\end{align}
Of course, in order for this definition to work, the quantity $\Tr \rho_A^n$ must be analytically continued to real $n$. This is always possible \cite{Calabrese:2004eu}: the eigenvalues $\lambda$ of $\rho_A$ satisfy $\Tr \rho_A=\sum_\lambda\lambda =1 $, and hence $\Tr \rho_A^n=\sum_\lambda\lambda^n$ is absolutely convergent and therefore analytic when Re $n > 1$. We can express this integral as
\begin{align}
    \Tr \rho_A^n = \frac{Z_n}{Z_1^n}
\end{align}
where
\begin{align}
   Z_n= \int_{\mathcal{M}_n(A)} [d\phi]\ e^{-S_E[\phi]}
\end{align}
The manifold $\mathcal{M}_n(A)$ consists of $n$ copies of the original manifold $\mathcal{M}$ with the sewing conditions
\begin{align}
    &\phi_k(\mathbf{x},0^+) = \phi_{k+1}(\mathbf{x},0^-) && \mathbf{x}\in A\nonumber \\
    & \phi_k(\mathbf{x},0^+) = \phi_{k}(\mathbf{x},0^-) &&\mathbf{x}\notin A
\end{align}
Note that $\mathcal{M}_1=\mathcal{M}$. Defining the dimensionless canonical free energies
\begin{align}
    \Fcan_n=-\log Z_n
\end{align}
we can express 
\begin{align}
    S_A = \partial_n\left(\Fcan_n-n\Fcan_1\right)\rvert_{n=1}
\end{align}
Let us now specialize to the $(2+1)$-dimensional case. We know that for a circular entangling surface $A$ of radius $R$ the entanglement entropy has a universal UV-divergence that is linear in $R$
\begin{align}
    S_{EE}(R)=\alpha\Lambda R+S_{\rm fin}(R)
\end{align}
Applying the defect filter ${\cal D}_{\rm def}=1-R\partial_R$, one obtains the entropic $F$-function of Ref.~\cite{CasiniHuerta:2012},
\begin{align}
\label{eq:entropicf}
		F(R)
		=
		-{\cal D}_{\rm def}\,
		\partial_n\big(\Fcan_n-n\Fcan_1\big)\big|_{n=1}.
\end{align}
Thus the entropic $F$-function is minus the $n\to1$ derivative of the replica-energy filter applied to the entangling defect.\par 
A closely related example appears in the line-defect $g$-theorem of Ref. \cite{Cuomo:2021rkm}. There, after subtracting the ambient bulk contribution, the circular defect free energy has a single local power-law ambiguity, namely the defect cosmological-constant term proportional to the length $R$. The corresponding scheme-independent quantity is therefore obtained by the same first-order Euler filter, $1-R\partial_R $.
This is the same kinematic mechanism that appears above: for an effectively one-dimensional defect, the only local extensive ambiguity is a perimeter term, and the natural filter which removes it is linear in $R\partial_R$.\par
We stress, however, that the existence of this first-order filter does not by itself imply monotonicity. The filter is a kinematic device which removes scheme-dependent local terms and defines a candidate finite quantity. Monotonicity requires additional dynamical or information-theoretic input. In the entropic $F$-theorem this input is strong subadditivity, while in the line-defect theorem of Ref.  \cite{Cuomo:2021rkm} it is supplied by the defect Ward identities together with reflection positivity.

\subsubsection{Discussion}
We now discuss the structural differences between \eqref{eq:FE_def} and \eqref{eq:entropicf}. The function $F_{\cal E}$ is defined by applying a second-order filter to $\Fcan_{S^3}(R)=-\log|Z_{S^3}(R)|$. This filter is second order because it must remove two UV-divergent counterterms, one purely extensive, proportional to $R^3$, and one subextensive, proportional to $R$. By contrast, the entropic $F$-function is obtained by applying the first-order filter to the $n$-derivative of the replica-defect free energy, with the minus sign in Eq.~\eqref{eq:entropicf}. The purely extensive bulk contribution has already been removed by the replica subtraction $\Fcan_n-n\Fcan_1$. This explains the lack of monotonicity of $F_{\cal E}$: the double-filter polynomial is sign-indefinite, and no general positivity principle controls the resulting higher logarithmic derivatives of $\Fcan_{S^3}(R)$.

\subsection{$(3+1)$-dimensional fixed points: $S^4$ free energy and the $a$-theorem}
\label{subsec:atheorem_precise}

In even spacetime dimensions the vacuum functional on a smooth compact background exhibits a logarithmic divergence governed by the Weyl anomaly  \cite{DeserSchwimmer:1993,Duff:1994,Vassilevich:2003}.
For a CFT on the round four-sphere $S^4_R$ one has schematically
\begin{align}
	\label{eq:S4_structure}
		\Fcan_{S^4}(R)=\sum_{k=0}^{1}\alpha_k\Big(\Lambda R\Big)^{4-2k} + A_0 \log \left(\Lambda R\right)+\Fcan_{\rm fin},
\end{align}
where $\Lambda$ is a UV cutoff and the power-law terms are removable by local counterterms. On the round sphere the Weyl-squared contribution vanishes, and the universal logarithmic coefficient is fixed by the Euler anomaly coefficient $a$. With the convention used below, $A_0=-4a$  \cite{Duff:1994,DeserSchwimmer:1993,Minahan:2021_deformed_spheres}.

Applying Euler-type filters to Eq.~\eqref{eq:S4_structure} removes the local power laws and isolates the universal anomaly coefficient. Writing $D\equiv R\partial_R$, on a smooth closed $S^4$ the local counterterm sector generates divergent power laws proportional to $R^4$ and $R^2$, while curvature-squared counterterms shift the finite constant part. The canonical ``power-law filter'' is therefore
\begin{align}
	&\mathcal{D}_{\rm pow}^{(4)}\equiv \Big(1-\frac12 D\Big)\Big(1-\frac14 D\Big)
\end{align}
Since the universal information sits in the logarithmic term rather than in an $O(1)$ constant, the natural 4D analogue of the 3D double filter is an anomaly extractor obtained by one additional $\log R$ derivative,
\begin{align}
	\label{eq:anomaly_extractor_4d}
		{\cal A}_{\cal E}(R)\equiv -D\,\mathcal{D}_{\rm pow}^{(4)}\,\Fcan_{S^4}(R).
\end{align}
At a conformal point, $\mathcal{D}_{\rm pow}^{(4)}$ removes the power divergences and $D$ picks out the logarithmic coefficient, yielding
\begin{align}
	\label{eq:a_from_replica}
		{\cal A}_{\cal E}(R)\Big|_{\rm CFT}=-A_0=4a,
\end{align}
where $a$ is the Euler anomaly coefficient on the round sphere \cite{Minahan:2021_deformed_spheres}. This emphasizes the key difference with odd dimensions: in 4D there is no analogue of extracting a universal constant by filtering, because the finite constant part of $\Fcan_{S^4}$ is scheme dependent since it can be shifted by finite local curvature counterterms, whereas the logarithmic coefficient is universal.

RG irreversibility in four dimensions (the $a$-theorem) implies the endpoint ordering
\begin{align}
	a_{\rm UV}\ge a_{\rm IR},
\end{align}
with a nonperturbative proof via the dilaton effective action and dispersion relations \cite{KomargodskiSchwimmer:2011}. As in $d=3$, this is an endpoint statement. Indeed, constructing a scheme-independent, fully monotone interpolating function from $\Fcan_{S^4}(R)$ away from fixed points requires additional input beyond the scaling filter.

We emphasize that the rigorous monotonicity statements here are fixed-point statements: $\mathcal{D}_L$ (or $\mathcal{D}_{\rm ren}^{(3)}$) provides a clean extraction of $c$ or $F$ at conformal points, and RG irreversibility implies an ordering between the extracted values at UV and IR endpoints. Constructing a fully monotone interpolating function away from fixed points generally requires additional input, see  \cite{Zamolodchikov:1986,CasiniHuerta:2012}.

\subsubsection{Weyl anomaly on $S^4$ and extraction of $a$}
Consider $S^4_R$ with round metric $g_{\mu\nu}(R)=R^2\hat g_{\mu\nu}$. A change of radius is a global Weyl rescaling, and the trace Ward identity gives the exact relation
\begin{align}
	\label{eq:Ward_R_variation}
		\frac{d}{d\log R}\,\Fcan_{S^4}(R)
		= \int_{S^4_R}\! d^4x\,\sqrt g\,\langle \Theta^\mu{}_\mu\rangle .
\end{align}
At a four-dimensional CFT fixed point, the trace is the Weyl anomaly\footnote{We are neglecting the Pontryagin counterterm $ \epsilon^{\mu\nu\rho\sigma}{R_{\mu\nu}}^{\alpha\beta}{R_{\alpha\beta}}^{\rho\sigma}$ because there is no known unitary field theory model that gives it as Weyl anomaly \cite{Nakayama:2013is}.}  \cite{DeserSchwimmer:1993,Duff:1994, Nakayama:2013is}
\begin{align}
	\label{eq:Weyl_anomaly_4d}
	\langle \Theta^\mu{}_\mu\rangle =\frac{1}{16\pi^2}\Big(c\,W_{\mu\nu\rho\sigma}^2 - a\,E_4 + a'\,\Box \mathcal R\Big),
\end{align}
where $a$ and $c$ are scheme-independent, while $a'$ can be shifted by local $R^2$ counterterms. On the round sphere $S^4$ one has $W_{\mu\nu\rho\sigma}^2=0$ (conformal flatness) and $\int_{S^4}\sqrt g\,\Box\mathcal R=0$ (total derivative on a compact manifold without boundary), hence
\begin{align}
	\int_{S^4_R}\! d^4x\,\sqrt g\,\langle \Theta^\mu{}_\mu\rangle
	= -\frac{a}{16\pi^2}\int_{S^4_R}\! d^4x\,\sqrt g\,E_4 .
\end{align}
Using $\int_{S^4}\sqrt g\,E_4 = 32\pi^2\,\chi(S^4)=64\pi^2$  \cite{EguchiGilkeyHanson:1980,Minahan:2021_deformed_spheres}, we obtain
\begin{align}
	\label{eq:dlogR_W_equals_a}
		\frac{d}{d\log R}\,\Fcan_{S^4}(R)\Big|_{\rm CFT} = -4a ,
\end{align}
where we adopt the convention in which a free scalar has $a>0$, so that the coefficient in $\Fcan=-\log Z$ is $-4a$ on the round sphere.
Integrating \eqref{eq:dlogR_W_equals_a} implies that at a CFT
\begin{align}
	\label{eq:WS4_structure_CFT}
	\Fcan_{S^4}(R)=&\alpha_0(\Lambda R)^4+\alpha_1(\Lambda R)^2 -4a\log(\Lambda R)\nonumber\\
	&\;+\;(\text{scheme-dependent const}) .
\end{align}
Therefore, unlike odd dimensions, in $d=4$ the universal CFT data is encoded in the logarithmic coefficient. In particular, defining $D\equiv R\partial_R$ and $\mathcal D^{(4)}_{\rm pow}$ as defined above
the combination
\begin{align}
	\label{eq:a_extractor}
		\mathcal A_{\cal E}(R)\equiv -D\,\mathcal D^{(4)}_{\rm pow}\,\Fcan_{S^4}(R)
\end{align}
removes the power divergences and differentiates away scheme-dependent constants, yielding at a fixed point
\begin{align}
	\label{eq:A_equals_a}
	\mathcal A_{\cal E}(R)\Big|_{\rm CFT} = 4a .
\end{align}

\subsubsection{Free scalar mode analysis in $d=4$: a different obstruction}

The 4d extractor ${\cal A}_{\cal E}$ has a structure that is qualitatively different from the 3d double filter.
To see this, consider the free conformally coupled scalar of mass $m$ on~$S^4_R$. The eigenvalues of $-\nabla^2+2/R^2+m^2$ (the conformally coupled scalar operator on $S^4_R$) are $\lambda_n=n(n+1)/R^2+m^2$ with degeneracy $d_n=n(n+1)(2n+1)/6$, for $n=1,2,\dots$ \cite{BirrellDavies:1982}. Writing $x\equiv mR$ and $a_n\equiv n(n+1)$, a direct computation shows\footnote{Since for any diffeomorphism-invariant regularization all the scheme dependence is encoded in counterterms annihilated by the filter, the regularization of the canonical free energy commutes with its differentiation with respect to the sphere radius $R$. See Ref. \cite{FEletter} for more details on this argument and an explicit calculation on $S^3_R$.}
\begin{align}
\label{eq:AE_4d_mode}
\mathcal{A}_{\cal E}(x)
&=
4a
-
x^6\sum_{n=1}^{\infty}
\frac{d_n}{(a_n+x^2)^3}.
\end{align}
For a single real conformally coupled scalar, $4a=1/90$.
The infinite series is convergent.
The flow derivative follows by differentiation:
\begin{align}
\label{eq:dAE_4d}
	\frac{d{\cal A}_{\cal E}}{d\log R}=-6\,x^6\sum_{n=1}^\infty \frac{d_n\,a_n}{(a_n+x^2)^4}.
\end{align}
Unlike in the 3d free-scalar example, where the filtered quantity eventually develops a positive large-$x$ tail, the zeta-regularized contribution in the present 4d mode decomposition is the constant $1/90$, whose $\log R$ derivative vanishes. Thus the sign of $d{\cal A}_{\cal E}/d\log R$ is controlled entirely by the manifestly negative convergent sum in Eq.~\eqref{eq:dAE_4d}.

When $x\to\infty$ the theory becomes trivial, and its vacuum energy diverges. Correspondingly the sum in Eq. \eqref{eq:AE_4d_mode} diverges quartically. Indeed, for modes with $n\lesssim x$ one has $a_n\ll x^2$, so that $x^6/(a_n+x^2)^3\simeq 1$ and the summand behaves as $d_n\sim n^3/3$. Hence
\begin{align}
	\sum_{n=1}^\infty d_n\,\frac{x^6}{(a_n+x^2)^3}
	\;\gtrsim\;
	\sum_{n\lesssim x} d_n
	\sim \frac{x^4}{12}\,,
\end{align}
	and therefore ${\cal A}_{\cal E}(x)\to -\infty$.

\subsection{Observations}
The pattern in spacetime dimensions $D=2,3,4$ can be summarized as follows.
\begin{itemize}
	\item In $D=2$ (Subsec. \ref{subsec:2d_CFT}): the replica energy filter $(1-L\partial_L)$ is first order, and monotonicity follows from spectral positivity (the $c$-theorem).
	\item In $D=3$ (see Ref. \cite{FEletter}): the double filter is second order. It produces a quantity $F_{\cal E}$ with the correct endpoint values ($F_{\rm UV}$ and $F_{\rm IR}$) but is not monotone along the flow, due to the sign-indefinite filter polynomial. The entropic $F$-function instead comes from minus the $n\to1$ derivative of the first-order replica-energy filter applied to the replica-defect free energy.
	\item In $D=4$ (Subsec. \ref{subsec:atheorem_precise}): the combined operator $-D_R\cdot\mathcal{D}_{\rm pow}^{(4)}$, where $D_R=R\partial_R$, removes power-law divergences and extracts the
		logarithmic coefficient at fixed points. The flow derivative of ${\cal A}_{\cal E}= -D_R\cdot\mathcal{D}_{\rm pow}^{(4)}\Fcan $ is negative.
\end{itemize}
In all cases beyond two spacetime dimensions, additional dynamical input---such as strong subadditivity for entanglement entropy~\cite{CasiniHuerta:2012} or spectral positivity in the dilaton effective action~\cite{KomargodskiSchwimmer:2011}---is required to promote endpoint inequalities to monotone interpolants.

\section{Conclusion and classification}
\label{sec:conclusion}

The main result of this paper is that violations of additivity and extensivity in quantum field theory can be organized by applying scaling filters to the dimensionless canonical free energy $\Fcan=\beta F_{\rm can}=-\log Z$. The central object is the isotropic replica-energy functional
\begin{align}
\beta{\cal E}(L,\beta)
=
\left(1-\frac{1}{d}L\partial_L\right)\Fcan(L,\beta).
\end{align}
By construction, this filter annihilates the leading bulk term $-\beta P(\beta)L^d$ and isolates the non-homogeneous finite-size content of the thermal free energy. It therefore provides a unified diagnostic of extensivity violations across systems whose microscopic mechanisms and thermodynamic behavior are otherwise qualitatively different. We also introduced anisotropic replica energies ${\cal E}_{i=1,\cdots,d}$ and their corresponding directional scaling filters, which diagnose violations of additivity along individual spatial directions.

We applied this diagnostic to several classes of systems exhibiting nonextensive or nonadditive behavior:
\begin{itemize}
\item Genuinely nonextensive systems, such as a non-relativistic field in a Newtonian potential, where nonextensivity is the symptom of an instability, in accord with Ruelle's criterion \cite{ruelle}.

\item $1+1$-dimensional relativistic quantum field theories, where the replica-energy filter extracts the finite universal contribution associated with Zamolodchikov's $c$-function.

\item $2+1$-dimensional gapped phases on nontrivial spatial manifolds, whose dimensionless canonical free energy contains robust subextensive $O(L^0)$ contributions depending on the topology of the spatial manifold.

\item Replica defects and the $2+1$-dimensional $F$-theorem. We clarified the relation between the bulk sphere quantity $F_{\cal E}(R)$ of Ref.~\cite{FEletter} and the entropic $F$-function of Ref.~\cite{CasiniHuerta:2012}. The former is obtained by applying a second-order scaling filter to the sphere free energy $\Fcan_{S^3}(R)=-\log |Z(S^3_R)|$. The latter is minus the $n\to1$ derivative of the first-order replica-energy filter applied to the replica-defect free energy:
\begin{align}
	\Delta\Fcan_n(R)&=\Fcan_n(R)-n\Fcan_1(R),\\
	F(R)
	&=
		-\left(1-R\partial_R\right)
		\partial_n\Delta\Fcan_n(R)\big|_{n=1}.
\end{align}
Its monotonicity is not a consequence of the bulk double filter, but of the additional positivity implied by strong subadditivity.

\item $3+1$-dimensional fracton phases, whose dimensionless canonical free energy contains subextensive contributions $O(L^s)$, with $0<s<d$.

\item $3+1$-dimensional free scalar fields on the four-sphere $S^4_R$. Local counterterms contribute to $\Fcan_{S^4}(R)=-\log Z(S^4_R)$ through powers $(\Lambda R)^4$, $(\Lambda R)^2$, and the logarithmic term $-4a\log(\Lambda R)$, where $a$ is the Euler anomaly coefficient. We introduced the canonical second-order power-law filter
\begin{align}
	\mathcal{D}_{\rm pow}^{(4)}
	=
	\left(1-\frac12 D\right)
	\left(1-\frac14 D\right),
	\qquad
	D=R\partial_R,
\end{align}
and, for a free massive scalar field, showed that
\begin{align}
	{\cal A}_{\cal E}(R)
	\equiv
		-D\,\mathcal{D}_{\rm pow}^{(4)} \Fcan_{S^4}(R)
\end{align}
is finite, monotonically decreasing as a function of $mR$, and equal to $4a$ at conformal fixed points.

\end{itemize}

A key lesson is that nonadditivity splits into sharply distinct regimes, which are cleanly organized by the large-$L$ scaling of ${\cal E}$. Subextensive violations, such as surface terms, topological constants, and fracton degeneracies, are compatible with the existence of the standard thermodynamic limit, $\Fcan/L^d\to-\beta P$, but are universally detected by ${\cal E}$. By contrast, genuine nonextensivity occurs when ${\cal E}/V$ fails to vanish as $L\to\infty$, so that the usual thermodynamic limit breaks down. In the field-theory language of Sec.~\ref{sec:gravFT}, a sufficient mechanism is the presence of a nonintegrable static, or zero-Matsubara, mediator coupled to an operator with nonzero one-point function, as encoded by the nonextensivity criterion and illustrated by unscreened Newtonian attraction.

We summarize the hierarchy in Table~\ref{tab:hierarchy}, with the particle-number term absent and isotropic linear size $L$.
\begin{widetext}
\begin{center}
\renewcommand*{\arraystretch}{1.3}
\begin{tabular}{|c|c|c|}
\hline
Regime & $\Fcan$ & $\beta{\cal E}$ \\
\hline
\hline
Finite-size, short-range &
$-\beta P L^d+\sigma L^{d-1}+\cdots$ &
$\frac{1}{d}\sigma L^{d-1}+\cdots$, $\beta{\cal E}/L^d\to0$ \\
\hline
Topological liquid &
$-\beta P L^d-\kappa+\cdots$ &
$-\kappa+\cdots$ \\
\hline
Fracton / foliated &
$-\beta P L^d-\gamma L^{s}+\cdots$, $0<s<d$ &
$-\left(1-\frac{s}{d}\right)\gamma L^{s}+\cdots$ \\
\hline
Genuine nonextensive &
Superextensive, e.g.\ $|\Fcan|\sim L^{d+\delta}$ &
$|\beta{\cal E}|\to\infty$ \\
\hline
\end{tabular}
\captionof{table}{Hierarchy of additivity and extensivity violations diagnosed by the replica-energy filter
$\beta{\cal E}=(1-\frac1dL\partial_L)\Fcan$ with the particle-number term absent, for isotropic linear size $L$.}
\label{tab:hierarchy}
\end{center}
\end{widetext}

Finally, replica-energy filters are kinematic projectors, not by themselves monotonicity theorems. They remove the locally extensive or power-law pieces of $\Fcan=-\log Z$ and expose the universal data left behind: central charges in two dimensions, sphere free energies and defect free energies in three dimensions, topology-dependent constants, fracton subextensive degeneracies, and anomaly coefficients in four dimensions. Whether the resulting quantity is a genuine RG monotone depends on additional dynamical input. In two spacetime dimensions, this input is the spectral positivity underlying the $c$-theorem. In three spacetime dimensions, the monotone entropic $F$-function is the signed $n\to1$ derivative of the defect replica-energy filter, and its monotonicity follows from strong subadditivity. In four spacetime dimensions, the analogous endpoint inequality is controlled by the positivity structure of the dilaton effective action. Replica energy therefore provides a common thermodynamic language for additivity, defect free energies, and RG irreversibility, while also explaining why different filtered quantities can share the same fixed-point data but differ away from criticality.

\bibliography{References}
\bibliographystyle{apsrev4-2}

\end{document}